\documentclass{aa}  
\usepackage{graphicx}
\usepackage{txfonts}
\begin{document}

   \title{Switchbacks as signatures of magnetic flux ropes generated by interchange reconnection in the corona}

   \subtitle{}

   \titlerunning{Switchbacks as signatures of flux ropes}

   \author{J. F. Drake \inst{1,2}
          \and O. Agapitov \inst{3}
          \and  M. Swisdak \inst{2}
          \and S. T. Badman \inst{3}
          \and S. D. Bale \inst{3}
          \and T. S. Horbury \inst{4}
          \and Justin C. Kasper \inst{5,6}
          \and R. J. MacDowall \inst{7}
          \and F. S. Mozer \inst{3}
          \and T. D. Phan \inst{3}
          \and M. Pulupa \inst{3}
          \and A. Szabo \inst{8}
          \and M. Velli \inst{9}
          }

   \institute{Department of Physics, the Institute for Physical Science and Technology and the Joint Space Institute, University of Maryland, College Park, MD\\ \email{drake@umd.edu}
     \and
     Institute for Research in Electronics and Applied Physics, University of Maryland, College Park, MD\\
     \and
     Space Sciences Laboratory, University of California Berkeley, Berkeley, CA, USA\\
     \and
     The Blackett Laboratory, Imperial College London, London, UK.\\
     \and
     BWX Technologies, Inc., Washington DC\\
     \and
     Climate and Space Sciences and Engineering, University of Michigan, Ann Arbor, MI\\
     \and
     Code 695 NASA Goddard Space Flight Center, Greenbelt, MD\\
     \and
     Code 672 NASA Goddard Space Flight Center, Greenbelt, MD\\
     \and
     Department of Earth, Planetary and Space Sciences, University of California, Los Angeles, CA
             }

   \date{}

 
  \abstract
 {The structure of magnetic flux ropes injected into the solar wind
during reconnection in the coronal atmosphere is explored with
particle-in-cell simulations and compared with in situ
measurements of magnetic ``switchbacks'' from the Parker Solar
Probe. We suggest that multi-x-line reconnection between open and
closed flux in the corona injects flux ropes into the solar wind
and that these flux ropes convect outward over long distances
before eroding due to reconnection. Simulations that explore the magnetic structure
of flux ropes in the solar wind reproduce the following key features of the
switchback observations: a rapid rotation of the radial magnetic
field into the transverse direction, which is a consequence of reconnection
with a strong guide field; and the potential to reverse the radial
field component. The potential implication of the injection of large
numbers of flux ropes in the coronal atmosphere for understanding the
generation of the solar wind is discussed.}

   \keywords{Sun:magnetic field, solar wind, magnetic reconnection }

   \maketitle
%

\section{Introduction}\label{intro}

   A major discovery of Parker Solar Probe (PSP) was observations of
   large numbers of localized radial velocity spikes and associated
   reversals or switchbacks in the local radial magnetic field near
   the first perihelion at 35.7$R_\odot$
   \citep{Kasper19,Bale19,DudokdeWit20,Horbury20,Krasnoselskikh20,Mozer20,Phan20}. In
   Fig.~\ref{overview}, we present an example of such a velocity
   enhancement and the associated magnetic structure. The results are
   expressed in heliospheric coordinates in the radial direction, the
   $T$ direction and the $N$ direction \citep{Bale16,Kasper16}.  The
   radial velocity increases sharply during the event. Such velocity
   enhancements had occasionally been observed, although with greatly
   reduced frequency, in the polar solar wind
   \citep{Balogh99,Yamauchi04}, at 1AU \citep{Gosling09,Gosling11} and
   at 0.3AU \citep{Horbury18}. The increase in radial velocity is
   accompanied by a sharp rotation of the magnetic field from the
   negative radial direction into the $N$ direction with the overall
   magnetic field amplitude remaining nearly constant. The sharp
   rotation with the magnetic field amplitude remaining nearly
   constant and the radial magnetic field changing sign is a typical
   characteristic of these events. The sharp rotation of ${\bf B}$
   into the $N$ rather than the $T$ direction is somewhat unusual and
   is discussed further, later in the paper.

A key question as a result of these observations is whether the
intrinsic structure of the solar wind and its drive mechanisms are
being revealed by these data. The systematic positive nature of the velocity
spikes eliminated magnetic reconnection in the local solar wind as a
source of these spikes since local reconnection would produce spikes
both toward and away from the Sun \citep{Phan20}. That the
switchbacks resulted from the spacecraft crossing of the heliospheric
current sheet was also eliminated because the direction of the
electron strahl with respect to the local magnetic field did not
reverse as the magnetic field reversed \citep{Kasper19}. Another key
characteristic of the first perihelion of PSP was the possible
magnetic connection of the spacecraft to a small coronal hole
\citep{Bale19}, which suggested that magnetic reconnection between
open and closed flux near the solar surface \citep{Fisk05,Fisk20}
might be the source of the velocity spikes and switchbacks. On the
other hand, it seems implausible that the kinked magnetic field from
reconnection deep in the corona could propagate large distances
outward without straightening into an unkinked state \citep{Wyper18}.

However, the traditional picture of magnetic reconnection taking place
at a single magnetic x-line has now been supplanted by the view that
the narrow current layers that develop during reconnection in weakly
collisional \citep{Biskamp86,Daughton09,Bhattacharjee09,Cassak09} or
collisionless plasma \citep{Drake06,Daughton11} form multiple flux
ropes in systems with an ambient guide field. A flux rope, in contrast
with a magnetic island, is a magnetic structure with a magnetic field
that wraps around a strong axial magnetic field. The observation that
the magnetic field rotates sharply away from the radial direction with
nearly constant magnitude eliminates switchbacks as magnetic islands,
which have no strong axial magnetic field. Thus, the important
question is not whether the magnetic kink from a single reconnection
site deep in the corona can propagate significant distances outward in
the solar wind without straightening but whether flux ropes can
maintain their integrity as they propagate outward from the Sun. In
Fig.~\ref{schematic}, we present a schematic of the magnetic geometry
expected for a flux rope propagating outward in the solar wind. We note
that the flux rope is sandwiched within a unidirectional magnetic
field and that the flux rope has a strong axial field in addition to
the in-plane magnetic flux shown in the diagram. A key point is that
the in-plane magnetic field on one side of the flux rope is
parallel to the ambient magnetic field but on the other side it is antiparallel. The schematic is drawn in the solar wind frame in
which the flux rope has a significant radial velocity. Thus, there is
a strong velocity shear across the region of reversed magnetic field
which can suppress reconnection \citep{Chen97} when the velocity shear
is below the Kelvin-Helmholtz instability threshold.

Models based on Alfv\'enic turbulence have also been proposed to
explain the switchbacks \citep{Landi06,Squire20,Tenerani20} and have
been motivated by the striking correlation between the time evolution
of the plasma velocity and Alfv\'en velocity in switchback
observations \citep{Kasper19,Phan20}. The radial expansion of the
solar magnetic field leads to the amplification of Alfv\'enic
structures \citep{Jokipii89}. The expanding box model of Alfv\'en
waves has established that even low amplitude Alfv\'en waves close to
the Sun evolve to a strongly turbulent state with local reversals in
the radial magnetic field with a nearly constant magnetic field
strength as seen in the data \citep{Squire20}. On the other hand, a
key observation is the sharp rise in the ion temperature at the
boundaries of the switchback \citep{Farrell20,Mozer20}. This must be
explained by any proposed switchback model. Magnetic reconnection is
known to increase the ion temperature \citep{Gosling07,Drake09}. It is
unclear how Alfv\'enic turbulence would produce and maintain such
temperature jumps.

In the present manuscript we focus on two key issues: whether
reconnection between open and closed flux low in the corona is
generically bursty and is therefore a prolific source of flux ropes;
and whether the magnetic structure of flux ropes in the solar wind
reproduces the magnetic structure of the switchback
measurements. Thus, we are not presenting a full birth to death model
of flux ropes injected into the solar wind but are establishing the
key components that would lead to a complete model. Finally, we
discuss the Alfv\'enic nature of the measured velocity and magnetic
structures and the potential of the injected flux ropes to contribute to
the overall solar wind drive.

\section{PIC model and initial conditions}
\label{picmodel}
We carry out two distinct 2-D particle-in-cell (PIC) simulations, one
that focuses on the reconnection between open and closed flux low in
the corona (interchange reconnection) \citep{Fisk05} and a second that
focuses on the structure of flux ropes in the solar wind as shown in
Fig.~\ref{schematic}. The simulations of the low corona are carried
out with the PIC model on the basis of the low density (and therefore low
collisionality) of the open flux region. However, the results should
be model independent (MHD or PIC) since flux ropes form during
reconnection in weakly collisional
\citep{Biskamp86,Daughton09,Bhattacharjee09,Cassak09} as well as
collisionless \citep{Drake06,Daughton11} systems.
The simulations are performed with the PIC code p3d \citep{Zeiler02}.

The intial state for the interchange reconnection simulation consists
of a band of vertical flux (field strength $B_0$ in the negative
radial direction) with a low plasma density ($0.1n_0$) and an adjacent region with higher density that is a cylindrical
equilibrium with magnetic flux $\psi$ given by
\begin{equation}
  \psi \propto e^{-r^2/a^2-r^4/a^4}
  \label{psi}
  \end{equation}
with the in-plane magnetic field given by $\hat{z}\times\nabla\psi$
and has a maximum value of $0.76B_0$. The density in the cylinder has
the same functional form as $\psi$ but with a floor of $0.1n_0$ such
that the peak density is $n_0$. The temperature is uniform with
$T_e=T_i=0.25m_iC_{A0}^2$ with $C_{A0}$ the Alfv\'en speed based on
$B_0$ and $n_0$. The guide field $B_z$ is nonzero everywhere except in
the region with vertical flux and is chosen to balance the
pressure and tension forces. The peak value of $B_z$ is $1.09B_0$ at
the center of the region of cylindrical flux. The vertical and cylindrical field
slightly overlap in the initial state and have opposite directions so
that reconnection quickly onsets. 

The simulation to produce the solar wind flux rope in Fig.~\ref{schematic}
consists of an ambient uniform magnetic field $B_0$ with a region of
reversed magnetic field $\delta B_0$ and associated guide field $B_z$
so that the total magnetic pressure is constant. The initial density
$n_0$ and temperatures $T_e=T_i=0.15m_iC_{A0}^2$ are uniform. The
specific form for the reconnecting field component $B_x(y)$ is given
by
\begin{equation}
  B_x(y)=1-\frac{1+\delta}{2}\tanh\left(\frac{y-0.35L_y}{w_1}\right)+\frac{1+\delta}{2}\tanh\left(\frac{y-0.65L_y}{w_2}\right).
    \label{Bx}
\end{equation} 
For this magnetic configuration there are two current layers centered at
$y/d_i=0.35L_y$ and $y=0.65L_y$. The periodicity of $B_y$ is ensured
by additional current layers outside of the domain
$0:L_y$. Neither this initial state nor that for the interchange
simulation are rigorous kinetic equilibria, especially for ions, but
neither displays unusual behavior at early time. The results of both
simulations are presented in normalized units: the magnetic field to
the magnetic field $B_0$, times to the inverse proton cyclotron
frequency, $\Omega_i^{-1}=m_ic/eB_{0}$, and lengths to the proton
inertial length $d_i=c_{A0}/\Omega_i$. The mass ratio $m_i/m_e=25$ is
artificial as is the velocity of light ($20C_{A0}$ for the interchange simulation and $15C_{A0}$ for the solar wind simulation), but as has been
established in earlier papers, the results are not sensitive to these
values \citep{Shay07}. Key scale lengths for the interchange
simulation are $a=14d_i$ with the domain $L_x\times L_y=81.92d_i\times
81.92d_i$ with grid scales $\Delta_x=\Delta_y=0.02d_i$ and around 400
particles per cell. For the solar wind simulation $\delta=0.2$,
$w_1=d_i$, $w_2=8d_i$, $L_x\times L_y=40.96d_i\times 40.96d_i$ with
$\Delta_x=\Delta_y=0.05d_i$ and around 100 particles per
cell. Reconnection begins from particle noise.

\section{Simulation results: Flux rope generation during interchange reconnection}
\label{interchange:simulations}
In the interchange simulation the small overlap between the vertical
magnetic field and the cylindrical flux bundle causes magnetic
reconnection to quickly initiate. The merging process leads to a
well-developed current layer that thins and spreads in the vertical
direction, leading to the formation of the flux rope as shown in
Fig.~\ref{interchangeisland}. Shown is the out-of-plane current
$J_{ez}$ with overlying magnetic field lines in the plane of
reconnection ($x-y$ plane). In (a) at $\Omega_it=70$ is the developing
current layer, in (b) at $\Omega_it=90$ is the formation of the flux
rope in the current layer, and in (c) at $\Omega_it=110$ is the
vertical propagation of the flux rope in the region of open flux.  In
(d), (e) and (f) the magnetic field, velocity components and electron
and ion temperatures are shown in the horizontal cut through the flux
rope shown by the green line in (c).  The vertical magnetic field
(red) reverses sign across the flux rope while, as expected for a flux
rope, the axial magnetic field (blue) increases sharply within the
flux rope. The total magnetic field (black) is relatively constant
across the flux rope but exhibits distinct dips on either edge of the
flux as is often seen in the switchback data
\citep{Bale19,Farrell20}. In (e) the vertical velocity in red
increases to around $0.7C_{A0}$ on average so that the flux rope is
being injected upward with high velocity as expected. There are also
high velocity flows $V_{iz}$ (in blue) due to the magnetic curvature
in this direction. The velocity is on average in the positive $z$
direction (the same direction as $B_z$), which is the dominant
direction of the magnetic curvature in the out-of-plane direction. In
(f) the electron and ion temperature profiles reveal sharp increases
within the flux rope. The increase in the ion temperature has also
been documented within switchbacks \citep{Farrell20,Mozer20}.

One of the important observational constraints in any model to explain
switchbacks is that the direction of the electron strahl with respect
to the local magnetic field remains unchanged as the radial magnetic
field reverses direction in the switchback \citep{Kasper19}. The
direction of the strahl expected from a flux rope model is consistent
with this observation. A reasonable assumption is that the strahl is
ejected upwards on open field lines on the right side of
Fig.~\ref{interchangeisland}. As the flux rope first forms in
Fig.~\ref{interchangeisland}(b) the strahl electrons circulate
counterclockwise in the island. This counterclockwise motion is
maintained as the flux rope is injected into the region of
unidirectional flux. Thus, on the left edge of the flux rope in
Fig.\ref{interchangeisland}{c} the strahl electrons would move
downward, opposite to the direction of the local ${\bf B}$. Thus, the
direction of the strahl with respect to the local magnetic field is
unchanged inside of the flux rope compared with the ambient solar
wind, consistent with observations. The observation of strahl within
the switchback requires that the flux rope
maintain its connection to the Sun as it propagates outward in the
solar wind.

The simulation therefore confirms that interchange reconnection in the
corona is a source of flux ropes that could be ejected with high
velocity into the solar wind.  An important open question, of course,
is whether these ejected flux ropes can propagate large distances in
the solar wind to the location of PSP. Specifically, as shown in the
schematic in Fig.~\ref{schematic} and is evident in
Fig.~\ref{interchangeisland}(e), one side of the flux rope has a sign
of $B_R$ that is opposite to that of the ambient field outside of the
flux rope. Thus, erosion of the flux rope due to reconnection with the
ambient field is possible unless reconnection is suppressed due either
to the velocity shear \citep{Chen97} or diamagnetic stabilization
\citep{Swisdak10,Phan10,Phan13}.

We also emphasize that in simulations that do not produce flux ropes
(such as in smaller domains than that shown in
Fig.~\ref{interchangeisland}) the kinked magnetic field produced
during reconnection quickly straightens, eliminating the reversal in
the radial magnetic field seen in the switchbacks. Thus, the generation of
flux ropes seems essential for interchange reconnection to produce the
switchback structures within the solar wind.
\section{Simulation results: The structure of flux ropes in the solar wind}
\label{switchback:simulations}
The magnetic configuration defined in Eq.~(\ref{Bx}) and the following
paragraph is designed to produce a flux rope similar to that shown in
Fig.~\ref{schematic} and to reproduce the magnetic structure of
switchbacks seen in the PSP data. The flux rope in
Fig.~\ref{interchangeisland} is not suitable for this comparison
because the magnetic field wrapping the flux rope is comparable to
that outside in the region where it is formed while in the switchbacks
the reversed radial field, and therefore the magnetic field wrapping
the flux tube, is often smaller than the ambient solar wind radial
field. The $x$, $y$ and $z$ directions of the simulation correspond to
the direction of the ambient solar wind magnetic field, the normal
direction of the initial current layer (the direction of
inhomogeneity) and the out-of-plane direction (direction of
homogeneity in a 2D system). The initial reversed magnetic field is
taken to be weak, $\sim 0.2B_0$, much smaller than the ambient radial
field $B_0$ while the guide field $B_z$ in the region where the
initial radial field reverses is of order $B_0$ since the initial
condition is force free with the total magnetic field a constant
across the region of reversed flux. With this initial configuration
the magnetic field that wraps the flux rope after reconnection (the
$B_x$ and $B_y$ components) is weak compared $B_z$ as in the
switchback observations.

The time sequence of reconnection in the configuration is shown in
three snapshots of $J_{ez}$ in the $x-y$ plane in
Fig.~\ref{switchback}. Magnetic reconnection starts from noise at the
narrow current layer peaked at $y=14.3d_i$ (Fig.~\ref{switchback}(a)
at $\Omega_it=36$). There is negligible reconnection at the wider
current layer at $y=26.6d_i$. Reconnection proceeds at the narrow
current layer as small flux ropes merge (Fig.~\ref{switchback}(b) at
$\Omega_it=180$) until the largest island reconnects all of the
reversed flux, forming a flux rope that is, as in the schematic,
bounded by the radial magnetic field of the ambient solar wind
(Fig.~\ref{switchback}(c) at $\Omega_it=292$). The in-plane magnetic
field lines are shown in Fig.~\ref{switchback}(d) at the time in
(c). The state shown in (c) and (d) is, of course, transient because
the two flux ropes present at this time will merge into a single flux
rope. Again, we emphasize that the simulation is designed to produce
the flux large rope shown in Fig.~\ref{switchback}(c) and (d) so its
structure can be compared with the magnetic signatures of switchbacks
in the PSP dataset. As discussed in
Sec.~\ref{interchange:simulations}, we suggest that flux ropes are
generated within the corona and ejected into the solar wind. Flux
ropes injected in the solar wind are likely to undergo mergers as they
propagate. We further discuss the dynamics of merging flux ropes in
Sec.~\ref{discussion}.

In comparing the stucture of the flux ropes in the observations with
that of our simulation we represent the data in a coordinate
representation that differs from the usual heliospheric $R$, $T$, $N$
system. We reverse the sign of the radial magnetic field to match that
of our ambient solar wind magnetic field by defining $R'=-R$. We then
carry out a minimum variance analysis of the data in the $N-T$ plane
to define new coordinates $N'$ and $T'$ with $N'$ being the minimum
variance direction and $T'$ being orthogonal to $R'$ and $N'$. Thus,
$R'$ corresponds to our $x$ direction, $T'$ to our $z$ direction and
$N'$ to our $y$ direction. We compare cuts of the magnetic field in
the $y$ direction in our simulation with the time sequence of the PSP
observations. The spacecraft trajectory is, of course, not fully
aligned with the $N'$ direction. On the other hand, the observations
of switchbacks suggested that they are highly elongated with their
scale length in the radial direction being much larger than in the
$N'$ direction \citep{Horbury18}. For this reason, the time sequence
of the spacecraft data is insensitive to the crossing angle of the
flux rope (we are not interested in timing the crossings). The
comparison of the spacecraft data with cuts in our $y$ direction
should be accurate unless the spacecraft trajectory is directly along
the axis of the flux rope ($T'$ direction) or in the radial ($R'$)
direction. This assumption is also consistent with the high azimuthal
velocity of PSP near the perihelion \citep{Bale19}. At times before
and after perihelion the trajectory of the spacecraft through the
switchbacks is likely to be much more complex but the high elongation
of the switchbacks should mitigate uncertainties about the angle of
the trajectory through the structure.

In Fig.~\ref{cuts}, we show two cuts
of our data and associated hodograms of the magnetic field along the
two white lines through the dominant flux rope in
Fig.~\ref{switchback}(c) and compare the results with the time
sequence and hodograms from two representative PSP switchback
events. In (c) and (g) are the time sequences of the magnetic field
components and magnitude while in (d) and (h) are the corresponding
hodograms. As reported in earlier papers, both events exhibit a sharp
rotation from the $R'$ direction (red) into the $T'$ (blue) direction,
with the $B_{R'}$ magnetic field taking on modestly negative values and
the total magnetic field magnitude being nearly constant
\citep{Kasper19,Bale19,Farrell20}. We note, however, the dip in the total
magnetic field at the edges of the switchback as seen in the flux rope
in Fig.~\ref{interchangeisland}(c). Such dips have been reported
previously \citep{Bale19,Agapitov20,Farrell20}. The hodogram maps a
nearly circular trajectory in the $R'-T'$ plane and swings quickly
from the $R'$ into the $T'$ direction where it remains for a
significant time before swinging back to the $R'$ direction. The two
switchback events are chosen to illustrate cases in which the minimum
variance magnetic field component (green) remains nearly zero (in
(c)) and takes on modestly negative values (in (g)). We compare with
data from cuts across the dominant flux rope in
Fig.~\ref{switchback}(c). In Fig.~\ref{cuts}(a) and (b) are the data
from the cut through the middle of the flux rope, where the magnetic
field $B_y$ within the flux rope is small, corresponding to the PSP
data in (c) and (d). The cuts through the simulation data are in
surprisingly good agreement with the observations. The magnetic field
rotates sharply from the $x$ to the $z$ direction, where it
remains before rotating sharply back to the $x$ direction. The
 magnetic field $B_x$ within the flux rope reverses over a portion of
the flux rope while the hodogram reveals that the magnitude of the
magnetic field is nearly constant. The data from the cut through the
region with negative $B_y$ is shown in (e) and (f). Again, the
magnetic structure of the flux rope matches well the satellite data
shown in (g) and (h).

The satellite data shown in Figs.~\ref{cuts}(c) and (g) reveal that
$B_{R'}$ (red) and $B_{N'}$ (green) within the switchback are highly
irregular with the axial, $B_{T'}$, magnetic field dominating. In our
interpretation of the data $B_{T'}$ is the axial field of the flux
rope while the two other components wrap
around the axial field to form the flux rope. In reconnection
observations in the Earth's magnetosphere and the solar wind at 1AU,
it has been a major challenge to establish the magnitude and even the
direction of the magnetic field that is normal to the reconnecting
current sheet. Thus, it is perhaps not surprising that directly
measuring the magnetic flux that wraps the flux rope is also a
challenge. The cuts through the flux rope in the simulation reveal
that the magnetic field $B_x$ within the flux rope changes from a
positive to a negative value across the flux tube. The positive and
negative values are small because the reversed field in the initial
state was taken to be small. Because $B_{R'}$ and $B_{N'}$ (red and
green) fields within the switchback data in Figs.~\ref{cuts}(c) and
(g) are small and irregular, identifying the expected reversal of the
magnetic flux is a challenge.

\begin{figure}
 \includegraphics[keepaspectratio,width=7.0in]{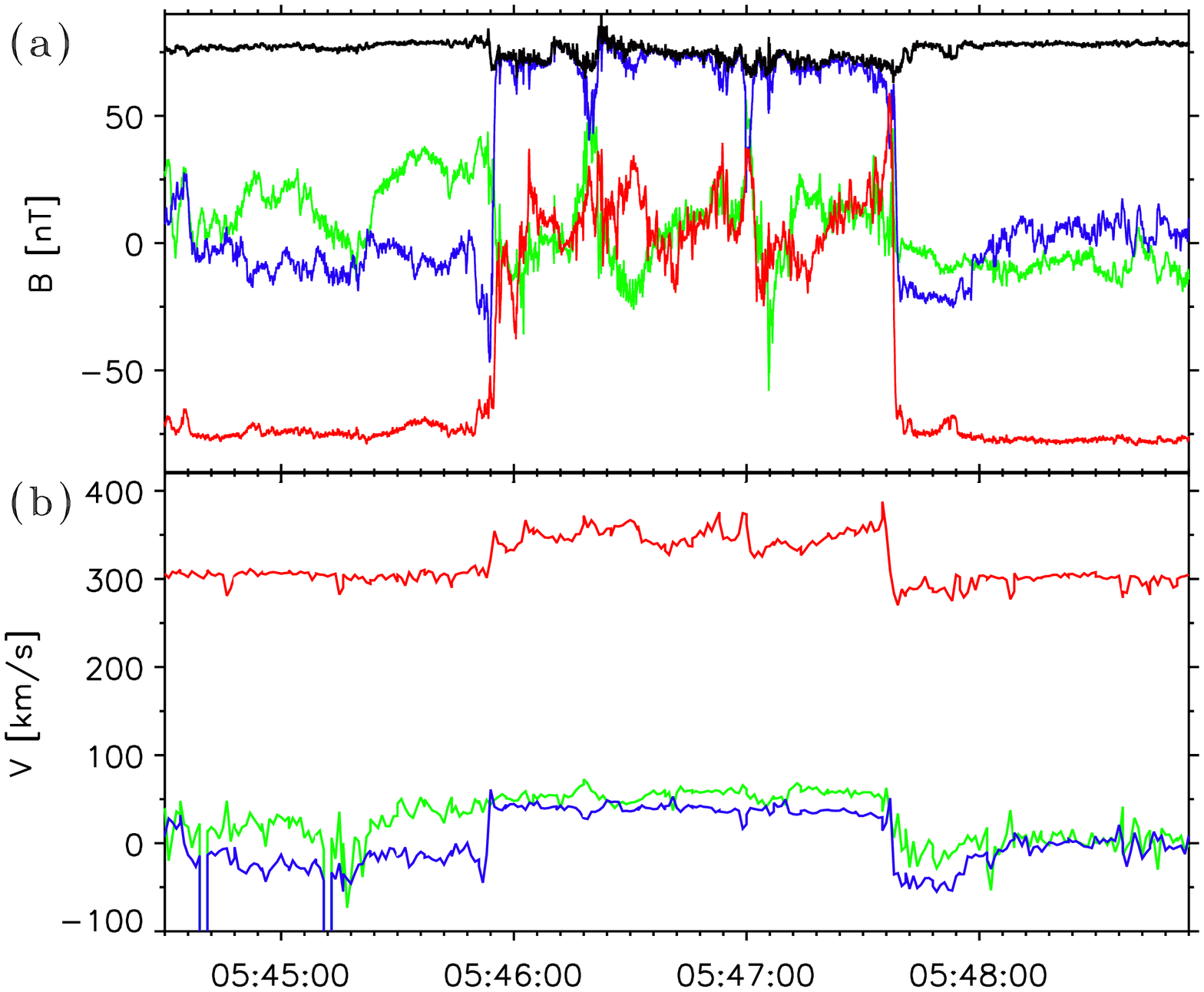} 
\caption{\label{overview} (Color online) From the PSP/FIELDS and PSP/SWEAP instruments on Nov.~5, 2018, measurements of the three components of the magnetic field and velocity in heliospheric $R$ (red), $T$ (green), $N$ (blue) coordinates at a time close to the first perihelion of the mission around 35.7$R_\odot$.}
\end{figure}

\begin{figure}
 \includegraphics[keepaspectratio,width=7.0in]{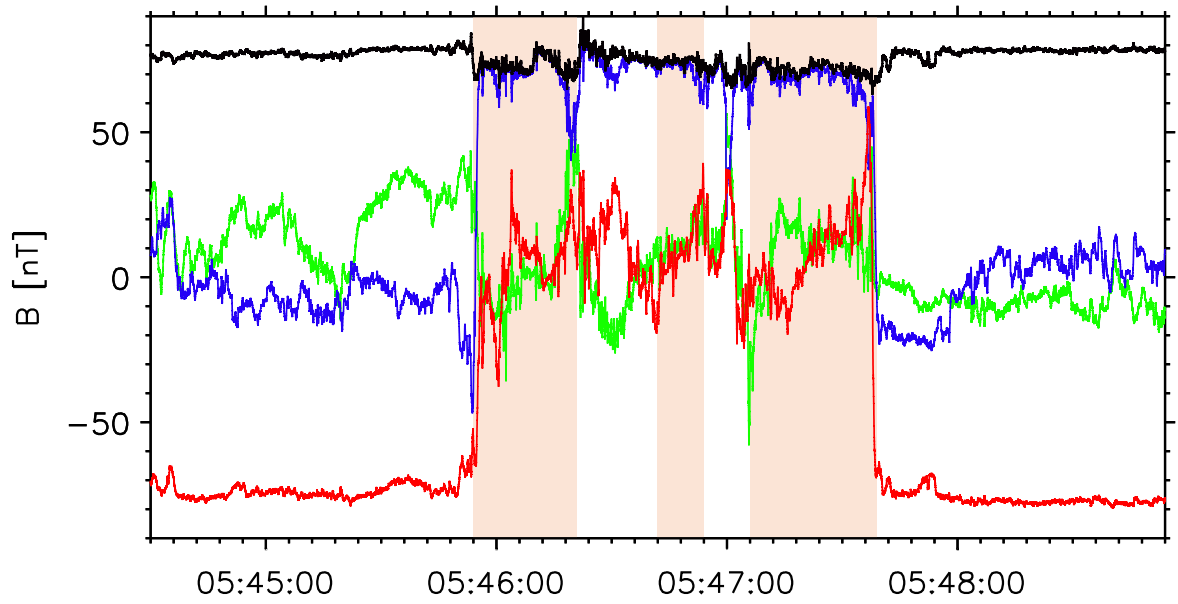} 
\caption{\label{ropes} (Color online) The same magnetic field data as in Fig.~\ref{overview}. The shaded regions mark the possible locations of three flux ropes embedded within a large switchback.}
\end{figure}

However, the event shown in Fig.~\ref{overview} displays large
variations in $B_R$ and $B_T$ within the switchback. First, we note that
$B_N$ actually exhibits three distinct peaks with clear dips around
05:46:20 and 05:47.00 separating those peaks. Within each peak the
radial magnetic field (in red) displays a distinct negative to
positive transition, as expected for a flux rope wrapped by a magnetic
field with components in the $R$ direction. Similar reversals in $B_T$
are evident. Thus, it appears possible that the switchback event in
Fig.~\ref{overview} corresponds to the crossing of three distinct flux
ropes. In Fig.~\ref{ropes}, we show the same PSP magnetic field data as
in Fig.~\ref{overview}. The shaded regions of Fig.~\ref{ropes} mark
the regions with the likely flux ropes. As stated earlier, this event
is unusual because of the large excursion in the $N$ direction. The
reason for our selection of this event is because of the large
velocity of the spacecraft in the $T$ direction. A switchback with a
large magnetic field component $B_T$ would mean that the spacecraft
trajectory is nearly aligned with the axis of the flux rope, making
the comparison with the simulation data a challenge.

Flux ropes have been studied in satellite data in other environments
so comparisons and contrasts with the PSP data can offer further
insight on switchback structure. In the magnetotail, for example, flux
ropes have been regularly documented \citep{Slavin03}. They have been
modeled as quasi-equilibrium structures that are magnetically
dominated so their structure is controlled by magnetic rather than
pressure forces. The magnetic field strength is typically peaked in
the magnetotail flux ropes since the gradient in the magnetic pressure
is balanced by the inward tension of the magnetic field that wraps the
flux rope. In the case of switchbacks, however, the magnetic field
that wraps the flux rope ($B_{N'}$ and $B_{R'}$ in
Figs.~\ref{cuts}(c), (g)) is small relative both to the solar wind
magnetic field outside of the flux rope, $B_{R'}$, and the axial
magnetic field within the flux rope, $B_{T'}$. The consequence is
that the tension force from the wrapping magnetic field is weak and
the magnetic field across the switchback is nearly constant.


\section{Discussion}\label{discussion}
\begin{figure}
  \includegraphics[keepaspectratio,width=6.0in]{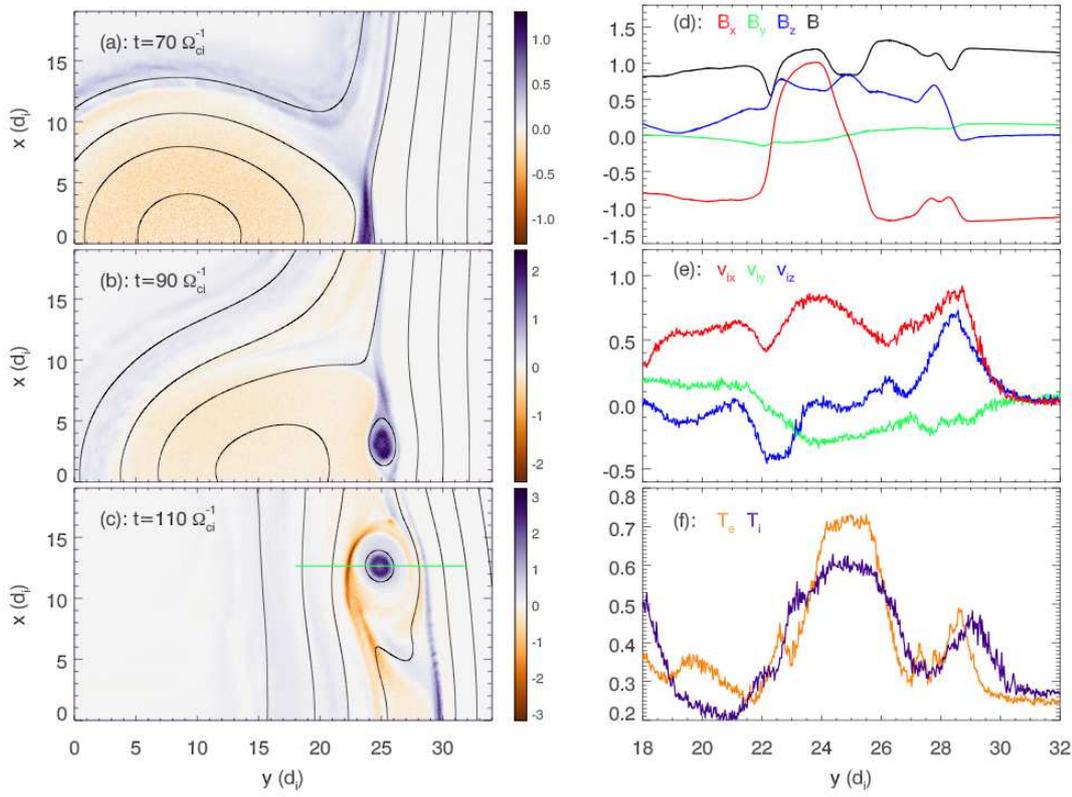}
\caption{\label{interchangeisland} (Color online) Flux rope formation
  at three times during interchange reconnection near the solar
  surface. In (a), (b) and (c) the out-of-plane electron current
  $J_{ez}$ with overlying magnetic field lines. Along the horizontal
  line in (c), in (d) the magnetic field components, $B_x$, $B_y$, $B_z$
  and $B$ in red, green, blue and black, respectively, in (e) the
  corresponding ion velocities and in (f) the electron and ion temperatures. }
\end{figure}

We have presented simulations of interchange reconnection between open
and closed flux in the low corona that reveal the formation of flux
ropes that are ejected with high velocity outward in the corona (see
Fig.~\ref{interchangeisland}). Cuts through the flux rope reveal that
a strong axial magnetic field is wrapped by magnetic flux and exhibit
the characteristic reversal in the radial magnetic field as documented
in switchback observations in the solar wind. The flux rope model
maintains the direction of the electron strahl with respect to the
local magnetic field as seen in the data.

The structure of flux ropes in the solar wind is explored with 2-D
reconnection simulations from an initial state with a band of reversed
radial magnetic flux sandwiched within a uniform solar wind magnetic
field. The magnetic structure of the resulting flux rope reveals
signatures that are consistent with switchback observations in the
solar wind: a sharp rotation of the ambient solar wind radial magnetic
field into the azimuthal direction; weak in-plane magnetic fields
within the structure with a local reversal of the radial magnetic
field component; and a nearly constant total magnetic field with
modest dips at the edges of the structure.

\begin{figure}
  \includegraphics[keepaspectratio,width=7.0in]{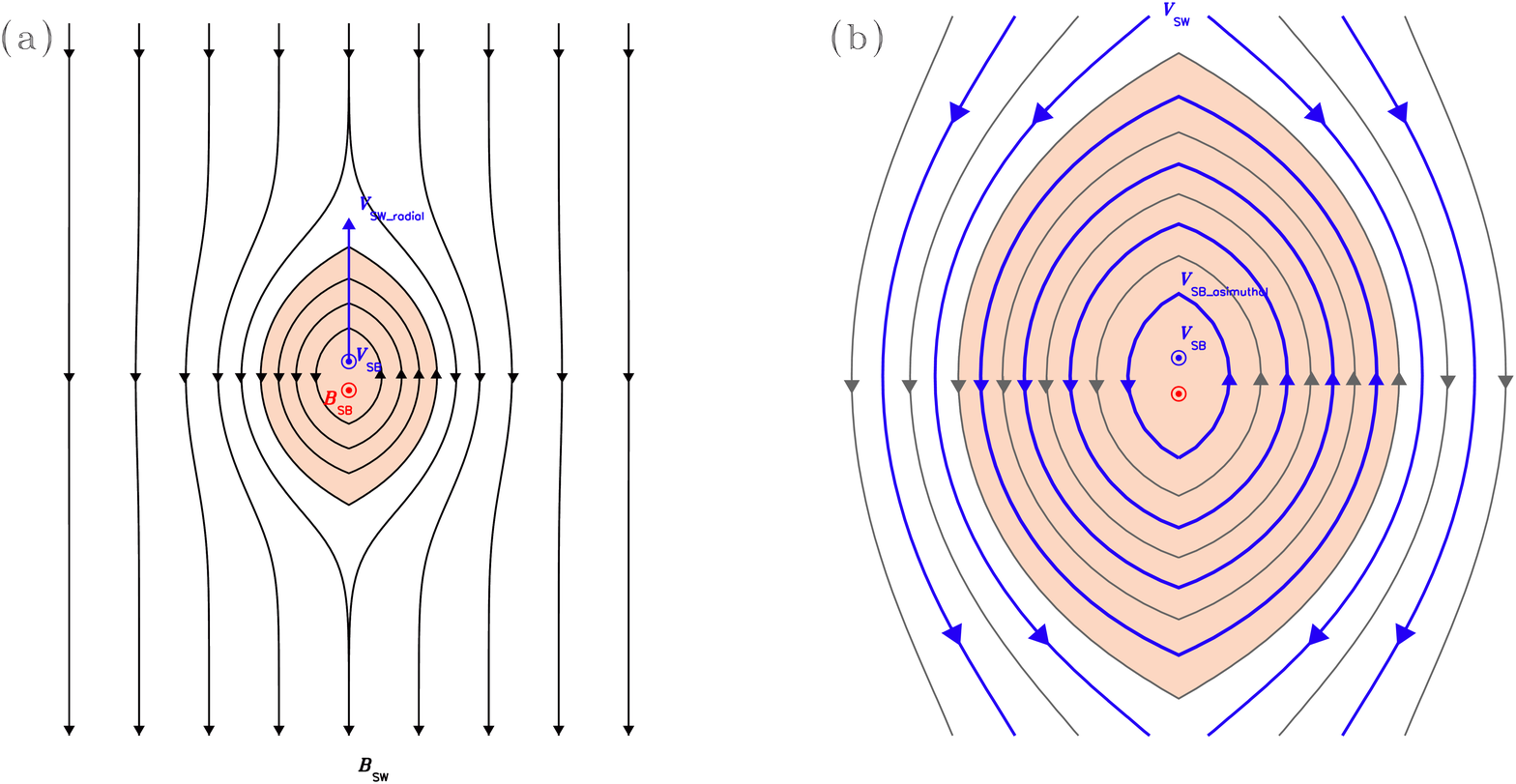}
\caption{\label{schematic} (Color online) In (a) a schematic of a flux
  rope with a magnetic field that wraps a core axial field. The flux
  rope is propagating outward within a solar wind magnetic field
  pointing back towards the sun. As indicated, the flux rope
  generally has axial flow. In (b) a schematic of the relaxed state
  of the flux rope in which the flow within the flux rope is field
  aligned. }
\end{figure}

\begin{figure}
  \includegraphics[keepaspectratio,width=6.0in]{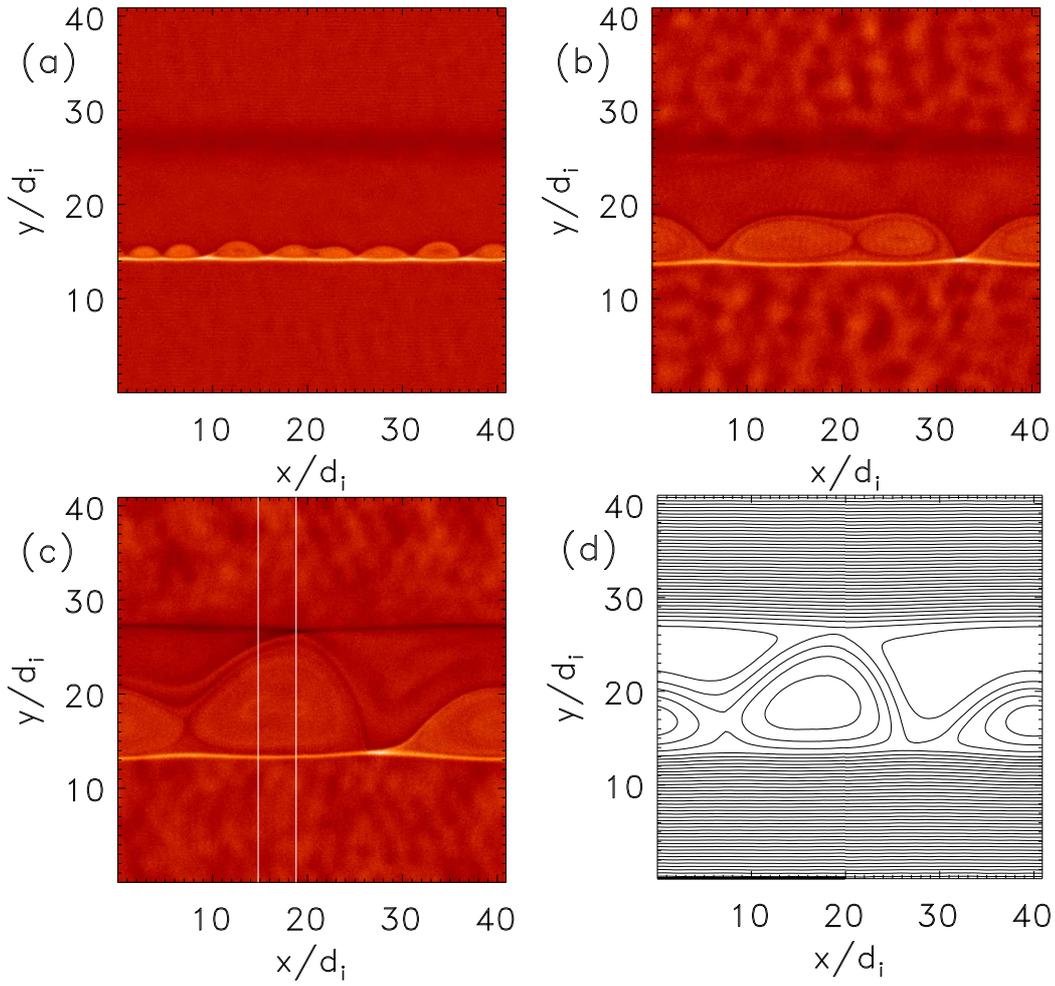}
\caption{\label{switchback} (Color online) The formation of the
  fluxrope within the ambient solar-wind magnetic field at
  $\Omega_it=$ 36 in (a), 180 in (b) and 292 in (c). Shown is the
  out-of-plane current $J_{ez}$. In (d) the magnetic field lines for
  the time in (c). The large flux rope in (d) has reconnected all of
  the initial reversed magnetic field $B_x$ and has the topology of the
  schematic in Fig.~\ref{schematic}. }
\end{figure}

\begin{figure}
  \includegraphics[keepaspectratio,width=6.0in]{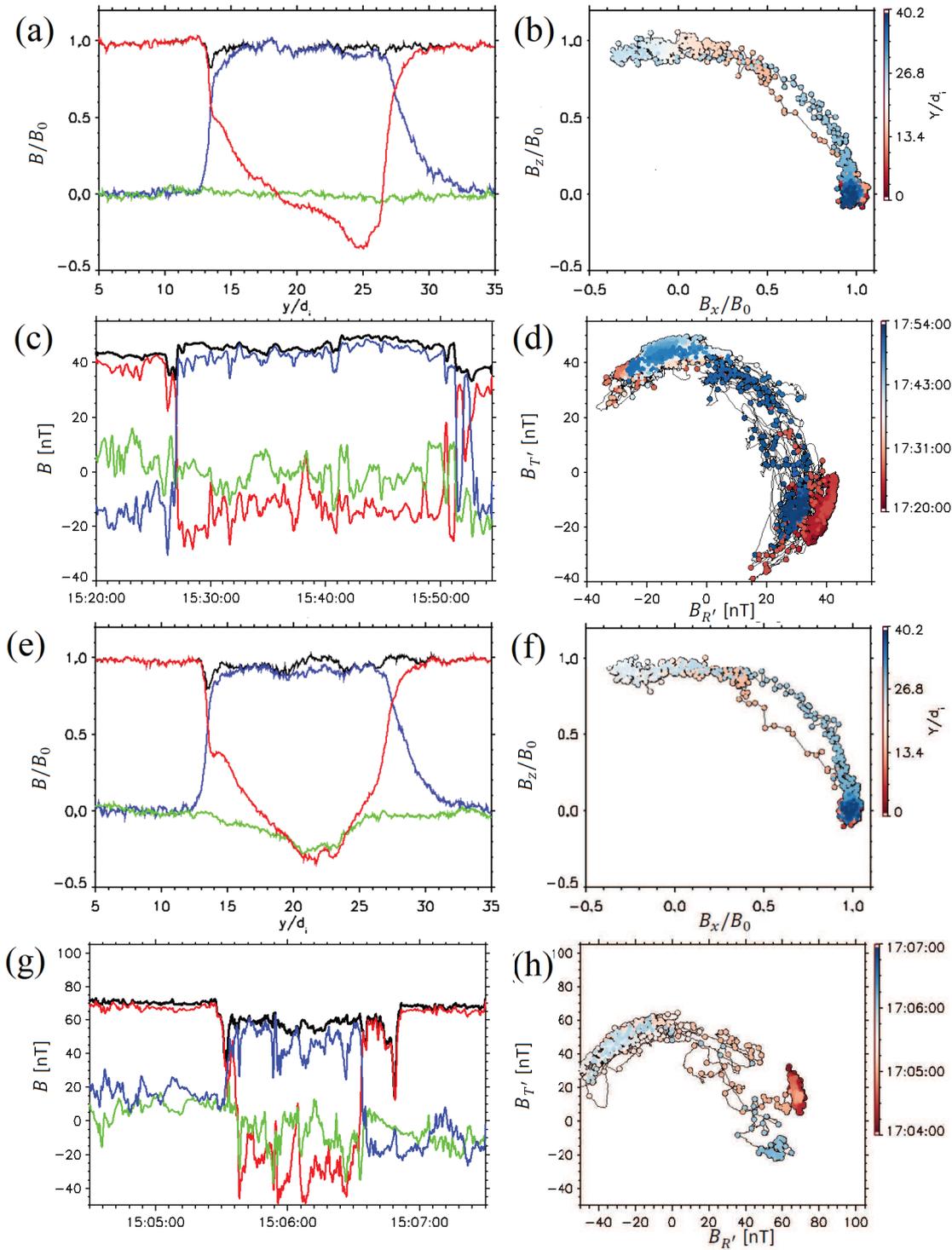}
\caption{\label{cuts} (Color online) In (a) and (e) cuts of the
  magnetic fields ($B_x$ in red, $B_y$ in green, $B_z$ in blue and $B$
  in black) and associated hodograms in (b) and (f) across the large
  flux rope in Fig.~\ref{switchback}. The cuts are along the white
  lines in (c). The cut in (a) corresponds to the midplane of the
  island where $B_y\sim 0$ while that in (e) is offset from the
  centerline where $B_y<0$. In (c) and (g) time profiles of magnetic
  fields ($B_{R'}$ in red, $B_{N'}$ in green, $B_{T'}$ in blue and $B$
  in black) and in (d) and (h) associated hodograms from switchbacks
  from PSP/FIELDS. The two events correspond to cases with the
  minimum variance magnetic field $B_{N'}$ small (Nov.~1, 2018)(as
  in (a)) and negative (Nov.~4, 2018) (as in (e)). See the text for a
  discussion of the coordinate system used to present the spacecraft
  data. It differs from the traditional $R$, $T$, $N$ system.}
\end{figure}

While the magnetic structure of the flux ropes in our simulations
display many of the characteristics seen in observations, they do not
display the striking proportionality between local flows, the magnetic
field \citep{Kasper19} and the Alfv\'en velocity \citep{Phan20}. We
suggest, however, that flux ropes ejected into the solar wind should
relax to a state in which the flows and magnetic field display the
Alfv\'enicity seen in the observations. There is a large literature on
the relaxation of flows in magnetized plasma systems
\citep{Hameiri83,Steinhauer98,Steinhauer99}. The general conclusion is
that flows relax to a state in which the flow is aligned with the
ambient magnetic field direction and are constant within a flux
surface. A bulk flow in an invariant direction that is constant on a
flux surface can also remain. The physics basis for this result seems
to be that perpendicular electric fields, which are required to
produce flows perpendicular to the magnetic field, tend to decay. In
contrast field aligned flows exist without an electric field. Thus, we
suggest that the outward flow of the flux rope in the schematic in
Fig.~\ref{schematic}(a) relaxes to that shown in
Fig.~\ref{schematic}(b), which is drawn in the frame of the flux
rope. The flow in this frame is along the local magnetic field and
includes out-of-plane flow, which is a general consequence of
reconnection with a guide field as shown in
Fig.~\ref{interchangeisland}(e) (discussion below). In the frame of
the flux rope, the flow of the solar wind is downward so in this
schematic the flow is everywhere aligned with ${\bf B}$ as in the
observational data. The confirmation that this relaxation takes place
in the solar wind will constitute an important extension of the
present work.

Another major surprise in the PSP dataset was the presence of a
transverse bulk flow of around 20$km/s$ in the heliospheric $T$
direction near perihelion \citep{Kasper19}. This flow was linked to
the strong positive values of $B_T$ during the rotation of the
magnetic field in the switchbacks. It has been suggested on the basis
of these observations that there is a general azimuthal ($T$
direction) circulation of magnetic flux and plasma flow as a result of
interchange reconnection in the low corona \citep{Fisk20}. What is
important to note in trying to interpret these observations is that
magnetic reconnection with a guide field drives strong field aligned
flows \citep{Lin93,Zhang19}. These field aligned flows are dominantly
in the out-of-plane direction, or in the direction of the axial
magnetic field in a flux rope, and scale like $V_z \sim \Delta
B_z/\sqrt{4\pi m_in}$, where $\Delta B_z$ is the characteristic
variation in the out-of-plane magnetic field across a reconnection
layer. They come about because in the presence of an out-of-plane
magnetic field the magnetic curvature ${\bf \kappa}={\bf b}\cdot
{\bf\nabla b}$ (with ${\bf b}={\bf B}/B$) has a component in the
out-of-plane direction. This means that interchange
reconnection with an ambient field component $B_T$ produces a
corresponding flow $V_T$ as seen in the data from the interchange reconnection
simulation in Fig.\ref{interchangeisland} and in the observational data. The $T$
directed momentum imparted to the plasma within the flux rope is, of
course, balanced by momentum transfer to the chromosphere. Thus, the
development of net flows in the $N-T$ plane should be expected in
regions where switchbacks exhibit a preferential direction.

Flux ropes in reconnecting current sheets typically first form at
small spatial scales as current sheets narrow
\citep{Biskamp86,Drake06,Bhattacharjee09,Cassak09}. Small flux ropes
then undergo mergers that lead to larger flux ropes. Large current layers
can produce a wide distribution of flux rope sizes
\citep{Fermo10,Fermo11}. Statistical models of the size distribution
of flux ropes suggest that the size distribution of large flux ropes
falls off exponentially and there is some observational support for
this behavior \citep{Fermo11}. Flux ropes injected into the solar wind
should undergo merging as they propagate away from the Sun. Normally
the magnetic islands that form during merging in a reconnecting
current layer become larger but their aspect ratio (of order unity)
does change since they expand into the region upstream of the current
layer as a result of their internal magnetic tension. However, in the
case of flux ropes propagating in a unidirectional magnetic field, the
merging process should lead to flux rope elongation. As revealed in
the data, the radial magnetic field within the switchback (the
magnetic field that wraps the flux rope) is typically smaller than
that of the ambient solar wind. This means that the tension force that
tries to make the flux rope round is much weaker than the
corresponding backwards acting tension force of the solar wind
magnetic field. As a consequence, the merged flux rope becomes
significantly longer and only modestly wider in the normal direction,
consistent with the high aspect ratio of the switchbacks measured in
the solar wind \citep{Horbury20}. The merging process also leads to
some reduction in the amplitude of the magnetic field wrapping the
flux rope while leaving the axial field relatively unchanged. Thus,
highly elongated flux ropes with weak wrapping magnetic fields might
be a consequence of flux rope merging as the structures propagate outward
in the solar wind.

Finally, the observations of substantial numbers of positive, radial
velocity spikes raises the question of the possible role of magnetic
reconnection in the corona as a direct drive of the solar wind outflow
from the sun. The local Alfv\'en speed in the low corona can be quite
large so that small outflows due to reconnection and the ejection of
flux ropes into the solar wind might be able to contribute to the
overall solar wind outflow. It is a question of how widespread
small-scale reconnection is in the corona. It is possible, for
example, that regions of open flux, where because of the low plasma
density the Alfv\'en speed is very high, release large numbers of
high-velocity flux ropes due to subsurface reconnection. As PSP moves
even closer to the outer reaches of the corona, the emergence of more
fine-scale structure of the solar wind would suggest that such a
hypothesis might be valid.

\begin{acknowledgements}
The authors acknowledge support from NSF grant No. PHY1805829, from
NASA grant NNX17AG27G, and from the FIELDS team of the Parker Solar
Probe (NASA Contract No. NNN06AA01C). O.A. and J.F.D. were supported
by NASA grant 80NNSC19K0848. O.A. was partially supported by NSF grant
No. 1914670. TSH was supported by UK STFC
ST/S000341/1. J.F.D. acknowledges partial support from NSF grant
No. PHY1748958 at the Kavli Institute for Theoretical Physics at
UCSB. The observational data used in this study are available at the
NASA Space Physics Data Facility (SPDF),
https://spdf.gsfc.nasa.gov/index.html. This research used resources of
the National Energy Research Scientific Computing Center, a DOE Office
of Science User Facility supported by the Office of Science of the
U.S. Department of Energy under Contract No. DE-AC02-
05CH11231. Simulation data is available upon request.
\end{acknowledgements}

%
%

\clearpage

\clearpage

\clearpage
\clearpage
\clearpage
\clearpage

\end{document}